\documentclass[twoside,10pt]{article}
\usepackage{amsmath,amssymb}
\usepackage{booktabs}
\usepackage{tam10}

\newcommand{\be}{\begin{equation}}
\newcommand{\ee}{\end{equation}}
\newcommand{\calo}{\mathcal{O}}

\title{\noindent A short introduction to Asymptotic Safety}
\shorttitle{Asymptotic safety}
\authors{Roberto Percacci$^1$\email{percacci@sissa.it}}
\shortauthors{R.\ Percacci}
\affiliations{
$^1$ SISSA, via Bonomea 265, I-34136 Trieste, and
\\
\phantom{$^1$} INFN, Sezione di Trieste, Italy
}

\abstract{
I discuss the notion of asymptotic safety and
possible applications to quantum field theories of gravity and matter.
}

\begin{document}
\maketitle

\section{What is asymptotic safety?}

We want to discuss the high energy behavior of a quantum field theory (QFT).
Assume that a ``theory space'' has been defined by giving
a set of fields, their symmetries and a class of action functionals
depending on fields $\phi$ and couplings $g_i$.
We will write $g_i=k^{d_i}\tilde g_i$, where $k$ is a momentum cutoff
and $d_i$ is the mass dimension of $g_i$.
The real numbers $\tilde g_i$ are taken as coordinates in theory space.
Ideally the couplings $g_i$ should be defined in terms of
physical observables such as cross sections and decay rates.
In any case ``redundant'' couplings, {\it i.e.} couplings
that can be eliminated by field redefinitions, should not be included.
We also assume that a Renormalization Group (RG) flow has been defined
on theory space; it describes the dependence of the action on an energy
scale $k$ (or perhaps a ``RG time'' $t=\log k$). 
The action is assumed to have the form
\begin{equation}
\label{Gammak}
\Gamma_{k}(\phi,g_i)=\sum_i g_i(k) \calo_i(\phi)\ ,
\end{equation}
where $\calo_i$ are typically local operators constructed with the field $\phi$ and
its derivatives, which are compatible with the symmetries of the theory.
We identify theories with RG trajectories.

It can generically be expected that when $k$ goes to infinity some couplings $g_i(k)$
also go to infinity. 
What we want to avoid is that the dimensionless couplings $\tilde g_i$ diverge.
In fact, there are famous examples such as QED and $\phi^4$ theory where this
happens even at some finite scale $k_{\rm max}$.
Such divergences signal a breakdown of the theory, and any theory where they occur
can only hold for a finite energy range, and is said to be an ``effective field theory''.
In contrast, suppose that the RG flow admits a fixed point (FP),
which is defined as a point $\tilde g_{i*}$ where the beta functions of the 
dimensionless couplings vanish.
An RG trajectory which ends (for $k\to\infty$) at the FP is free of such divergences;
it is called a ``renormalizable'' or ``asymptotically safe'' (AS)
trajectory and represents a UV complete theory \cite{wilson}.
The existence of such a trajectory is therefore a sufficient condition for the 
theory to be well behaved in the UV.

Now, let us try to count how many such trajectories there are in theory space.
We define the ``UV critical surface'' associated to our FP to be the subset
in theory space which is attracted towards it in the UV.
Assuming that this surface is a smooth manifold, its dimension is equal to the
dimension of its tangent space at the FP.
The latter can be computed in the following way.
Let $y_i=g_i-g_{i*}$; then in the vicinity of the FP the flow can be linearized:
\begin{equation}
\frac{{\rm d}y_i}{{\rm d}t}=M_{ij}y_j\ ,
\end{equation}
where
\begin{equation}
M_{ij}=\frac{\partial\beta_i}{\partial\tilde g_j}\Bigr|_{*} .
\end{equation}
By a linear transformation $z_i=S_{ij}y_j$ we pass to coordinates in which
$M$ is diagonal. Then the equation becomes
\begin{equation}
\frac{{\rm d}z_i}{{\rm d}t}=\lambda_i z_i\ ,
\end{equation}
where $\lambda_i$ are the eigenvalues of $M$.
The solutions of this equation are $z_i(t)=e^{\lambda_i t}z_i(0)$,
so the coordinates $z_i$ for which $\lambda_i<0$ are attracted towards the FP;
they are called the ``relevant'' couplings.
The coordinates for which $\lambda_i>0$ are repelled and are called ``irrelevant''.
If an eigenvalue vanishes the corresponding coordinate is said to be ``marginal''
and its behavior cannot be determined by the linearized analysis.
We will not consider such cases in the following, because they are not generic.
The conclusion then is that the dimension of the UV critical surface is equal to
the number of negative eigenvalues of $M$.

The condition of asymptotic safety requires that the theory has to lie in the
UV critical surface of the FP. This leaves a number of free parameters that is
equal to the dimension of this surface.
Thus, the theory is more predictive when the critical surface has lower dimension.
The ideal situation would be a theory with a one dimensional critical surface.
In this case there would be a single renormalizable trajectory and once we have determined
the initial position at some scale $k$, the theory is completely determined.
At the opposite extreme, if the UV critical surface was infinite dimensional, 
the theory would not be predictive.
The intermediate case is a theory space with finite dimensional critical surface.
Such a theory space would have the same good properties of a perturbatively renormalizable
and asymptotically free theory, because it would be well behaved in the UV 
and it would have only a finite number of undetermined parameters.

It is useful to consider the example of the Gaussian FP, 
which corresponds to a free theory. 
The beta functions have the form 
\begin{equation}
\frac{{\rm d}\tilde g_i}{{\rm d}t}=-d_i \tilde g_i+k^{-d_i}\beta_i\ .
\end{equation}
The functions $\beta_i={\rm d}g_i/{\rm d}t$ represent the loop 
corrections,
which vanish at the Gaussian FP. In this case the eigenvalues
of the matrix $M$ are given just by the canonical dimensions:
\begin{equation}
\lambda_i=-d_i\ .
\end{equation}
The relevant couplings are the ones that are power counting renormalizable,
and the critical surface consists of the power counting renormalizable actions.
We see that the requirement of asymptotic safety is a generalization
of the requirement of asymptotic freedom and renormalizability to the case
when the FP does not correspond simply to a free theory.
Of course the case of a non-Gaussian FP is harder to study.
If it is not too far from the Gaussian FP, one may be able to study it
using perturbation theory, but unlike asymptotically free theories,
in this case perturbation theory does not get better and better as the energy increases.

%%%%%%%%%%%%%%%%%%%%%%%%%
\section{Gravity}
%%%%%%%%%%%%%%%%%%%%%%%%%

Gravity is the domain of fundamental physics where the problem of 
finding a UV completion is most acute, and so it is here that most 
work on asymptotic safety has concentrated, following
the original suggestion of \cite{weinberg2}.
(For earlier reviews see \cite{reviews}.)
I will now show that it is reasonable
to expect that there exist asymptotically safe theories of gravity
\footnote{a complementary approach to the one discussed here consists in
performing Monte Carlo simulations of discretized gravity.
Significant advances have been made in recent years, also lending
support to the general idea of nonperturbative renormalizability.
See \cite{ambjorn} and references therein.}.

It is well known that general relativity
can be treated as an effective quantum field theory
\cite{Burgess,Espriu}.
This means that it is possible to compute quantum effects
due to graviton loops, as long as the momenta of the particles in
the loops are cut off at some scale.
For example, in this way it has been possible to unambiguously compute
quantum corrections to the Newtonian potential \cite{Donoghue}.
The results are independent of the structure of
any ``ultraviolet completion'', and therefore constitute genuine low
energy predictions of any quantum theory of gravity.
When one tries to push this effective field theory to energy scales
comparable to the Planck scale, or beyond, well-known
difficulties appear. It is convenient to distinguish two orders of
problems. The first is that the strength of the gravitational
coupling grows without bound. For a particle with energy $p$ the
effective strength of the gravitational coupling is measured by the
dimensionless number $\sqrt{\tilde G}$, with $\tilde G=G p^2$.
This is because the gravitational couplings involve derivatives of the metric.
The consequence of this is that if we let $p\to\infty$, also $\tilde G$
grows without bound. The second problem is the need of introducing
new counterterms at each order of perturbation theory. Since each
counterterm has to be fixed by an experiment, the ability of the
theory to predict the outcome of experiments is severely limited.

As we have seen in the previous section,
the first problem could be fixed if $\tilde G$ had a FP.
In order to see whether this is reasonable, imagine evaluating the
beta function using perturbation theory at one loop.
The coefficient\footnote{we choose units
such that $c=1$ and $\hbar=1$. Then everything has dimension of
a power of mass.} of the Hilbert action is the square of Planck's mass,
$M_\text{pl}^2=1/16\pi G$.
In the quantum theory it is expected to diverge quadratically
with the cutoff, leading to a beta function of the form
\begin{equation}
\label{planckrg}
k\frac{{\rm d}}{{\rm d}k}M_\text{pl}^2=c k^2\ ,
\end{equation}
where $c$ is some constant. 
Then, the beta function of $G$ has the form
$$
k\frac{{\rm d} G}{{\rm d} k}=-16\pi c G^2 k^2
$$
and the beta function of $\tilde G$ is
\begin{equation}
k\frac{{\rm d} \tilde G}{{\rm d} k}=2\tilde G-16\pi c \tilde G^2\ .
\end{equation}
This beta function has an IR attractive fixed point at $\tilde G=0$
and also an UV attractive nontrivial fixed point at $\tilde G_*=1/8\pi c$.
In order to establish whether $c>0$ one has to do a calculation.
The dependence of $G$ on distance has been computed at one loop
in the low energy effective field theory \cite{Bjerrum}, leading to
$$
16\pi c=\frac{167}{15\pi}\ .
$$
This has the desired positive sign, but it is not a particularly memorable number: 
it depends on details of the way in which it is computed.
Fortunately, one can show that for any reasonable cutoff it will
always have the same sign, so if one loop perturbation theory is a good guide, 
$\tilde G$ would indeed cease to grow at high energy and settle 
at some constant value of order one.

Of course such a value of $\tilde G$ is quite large and it is not really clear
that near this FP perturbation theory can be trusted. 
Furthermore, it is also known \cite{thooft}
that loop effects will induce terms with higher derivatives.
So the next thing one could do is calculate the one loop beta functions in 
a theory containing four derivative terms, with an action of the general form
\begin{equation}
\label{actionansatz}
\int {\rm d}^4x\,\sqrt{g}\left[2 Z\Lambda-Z R
+\frac{1}{2\lambda}C^{2}+\frac{1}{\xi}R^{2}+\frac{1}{\rho}E%+\frac{1}{\tau}\nabla^{2}R
\right]\ ,
\end{equation}
where $C^2$ is the square of the Weyl tensor, $E$ the integrand
of the Euler term,
$$Z=\frac{1}{16\pi G}\ ;\qquad \frac{1}{\xi}=-\frac{\omega}{3\lambda}\ ;\qquad 
\frac{1}{\rho}=\frac{\theta}{\lambda}\ .$$
Such calculations have a long history \cite{AvramidiBarvinski}.
They were mostly based on dimensional regularization.
More recently, we have repeated this calculation using a mass-dependent
heat kernel regularization procedure \cite{codello1}.
The beta functions of the four-derivative terms are
\begin{align}
\beta_{\lambda} & =  -\frac{1}{(4\pi)^{2}}\frac{133}{10}\lambda^{2}\ ;\nonumber\\
\beta_{\xi} & =  -\frac{1}{(4\pi)^{2}}\left(10\lambda^2-5\lambda\xi+\frac{5}{36}\right)\ ;\nonumber\\
\beta_{\rho} & =  \frac{1}{(4\pi)^{2}}\frac{196}{45}\rho^2\lambda\ .\nonumber
\end{align}
We see that the overall coupling $\lambda$ is asymptotically free:
\be
\lambda(k)=\frac{\lambda_0}{1+\lambda_0\frac{1}{(4\pi)^{2}}\frac{133}{10}\log\left(\frac{k}{k_0}\right)}\ ,
\ee
whereas the $\omega$ and $\theta$, which define the ratio of $\xi$ and $\rho$ to $\lambda$
tend to the asymptotic limits $\omega(k)\to \omega_*\approx -0.0228$ and
$\theta(k)\to \theta_*\approx 0.327$.
On the other hand, the cosmological constant and Newton's constant have the beta functions
\begin{align}
\beta_{\tilde \Lambda} & =
-2\tilde\Lambda
+\frac{1}{(4\pi)^{2}}\left[
\frac{1+20\omega^2}{256\pi\tilde G\omega^2}\lambda^2
+\frac{1+86\omega+40\omega^2}{12\omega}\lambda\tilde\Lambda\right]\nonumber\\
&\qquad\qquad\qquad\qquad\qquad-\frac{1+10\omega^2}{64\pi^2\omega}\lambda
+\frac{2\tilde G}{\pi}
-q(\omega)\tilde G \tilde\Lambda\ ,\\
\beta_{\tilde G} & =  2\tilde G
-\frac{1}{(4\pi)^{2}}\frac{3+26\omega-40\omega^2}{12\omega}\lambda\tilde G
-q(\omega) \tilde G^2\ ,
\end{align}
where $q(\omega)=(83+70\omega+8\omega^2)/18\pi$.
The first few terms in these expressions agree with \cite{AvramidiBarvinski},
but the last three terms of $\beta_{\tilde \Lambda}$ 
and the last term of $\beta_{\tilde G}$ are new.
The flow in the invariant subspace $\lambda=0$, $\omega=\omega_*$, $\theta=\theta_*$ is
\begin{align}
\beta_{\tilde \Lambda} & =
-2\tilde\Lambda
+\frac{2\tilde G}{\pi}
-q_*\tilde G \tilde\Lambda\ ,\\
\beta_{\tilde G} & =  2\tilde G-q_* \tilde G^2\ ,
\end{align}
where $q_*=q(\omega_*)\approx 1.440$.
This flow admits a FP with
$$
\tilde{\Lambda}_{*}=\frac{1}{\pi q_*}\approx 0.221\ ,\ \ \ \ \
\tilde{G}_{*}=\frac{2}{q_*}\approx 1.389\ .  
$$
It is quite striking that in spite of the very different structure of the theory,
the beta function of Newton's constant is very similar to the one we found in
Einstein's theory.
Again, the FP for $\tilde G$ occurs at some value of order one.
Nevertheless, it has been argued in \cite{niedermaier} that since $\lambda$,
the true coupling constant in this theory, is asymptotically free, 
this result is reliable.

These calculations highlight the importance of using a mass dependent cutoff scheme: 
had we used dimensional regularization, we would not see the nontrivial FP.
This is because dimensional regularization misses information about the power divergences.
It is therefore not a convenient method to study the beta functions of dimensionful couplings.

In fact, even with dimensional regularization there is a somewhat roundabout 
way to see the effect of power divergences:
they appear as logarithmic divergences in other dimensions.
One can therefore recover this information by performing a dimensional continuation.
In two dimensions $G$ is dimensionless and its beta function 
can be extracted at one loop from the pole of a counterterm.
It is $-38 G^2/3$ \cite{Kawai}.
Then, one can perform the so-called $\epsilon$ expansion, by studying the
beta function as a function of the dimension $d$.
For $d=2+\epsilon$, $G$ has dimension $\epsilon$, so $\tilde G=G k^\epsilon$.
The first term in the $\epsilon$ expansion gives
\begin{equation}
\beta_{\tilde G}=\epsilon\tilde G-\frac{38}{3}\tilde G^2\ ,
\end{equation}
so we recover the existence of a nontrivial FP in dimension $d>2$.
If we let $\epsilon=2$ the FP occurs again at some positive value $\tilde G$.
This was historically the first hint of asymptotic safety \cite{weinberg2}.

Both the one loop and the $\epsilon$ expansion give a FP which occurs
in a regime where the approximation is not clearly reliable.
It is for this reason that much of the recent work has been done using 
(some approximation to) an Exact RG Equation (ERGE),
which has been first applied to gravity in \cite{reuter1,dou}.
Without entering into details, suffice it to say that one can define 
a $k$-dependent effective action $\Gamma_k$
by introducing an IR cutoff $k$ in the functional integral, 
and that this functional obeys the equation
\begin{equation}
\label{ERGE}
k\frac{{\rm d}\Gamma_k}{{\rm d}k}=
\frac{1}{2}\mathrm{Tr}\left[
\frac{\delta^2\Gamma_k}{\delta\phi\delta\phi}+R_{k}\right]^{-1}k\frac{{\rm 
d}R_k}{{\rm d}k}\ .
\end{equation}
If $\Gamma_k$ has the form (\ref{Gammak}), 
\begin{equation}
k\frac{{\rm d}\Gamma_k}{{\rm d}k}=\sum_i \beta_i \calo_i(\phi)\ .
\end{equation}
Therefore, expanding the r.h.s. of (\ref{ERGE}) on the basis of operators ${\cal O}_i$
one can read off the beta functions of the individual couplings $g_i$.
This method has several advantages:
(i) it works in any dimension, (ii) there is no need to introduce UV regulators,
since the r.h.s. of (\ref{ERGE}) is finite, and 
(iii) it does not depend on the couplings being small.
Of course, it is generally impossible to compute the beta functions of infinitely many couplings and so one has to truncate the sum to finitely many terms.
For example, if we keep only the first two terms in (\ref{actionansatz}) we find,
for a cutoff of type ``Ib'' \cite{cpr2}:
\begin{equation*}
\begin{aligned}
\beta_{\tilde \Lambda}=&
\frac{-2(1-2\tilde\Lambda)^2\tilde\Lambda
+\frac{36-41\tilde\Lambda+42\tilde\Lambda^2-600\tilde\Lambda^3}{72\pi}\tilde G
+\frac{467-572\tilde\Lambda}{288\pi^2}\tilde G^2}{(1-2\tilde\Lambda)^2-\frac{29-9\tilde\Lambda}{72\pi}\tilde G}\ ,
\\
\beta_{\tilde G}=&  
\frac{2(1-2\tilde\Lambda)^2\tilde G
-\frac{373-654\tilde\Lambda+600\tilde\Lambda^2}{72\pi}\tilde G^2}
{(1-2\tilde\Lambda)^2-\frac{29-9\tilde\Lambda}{72\pi}\tilde G}\ .
\end{aligned}
\end{equation*}
One can still glean the one loop result, which is
obtained by neglecting $\Lambda$ and setting the denominators to one.
There has been a number of independent calculations, using different cutoffs
and different gauges, and treating the ghosts in different ways,
which give slightly different numbers but
agree on the qualitative structure of the result 
\cite{souma,lauscher,saueressig,litim,cpr2,eichhorn}.
This method has been applied also to four-derivative gravity in \cite{bms1},
where a nontrivial FP with nonzero values for all the couplings is found.

In another direction, it has been possible to work out the beta functions
for truncations of the form 
\be
\Gamma_k=\sum_{i=0}^n g_i \int {\rm d}^4x\sqrt{g}R^i\ .
\ee
The case $n=2$ was first examined in \cite{reuter2}, while in \cite{cpr1,ms}
the calculation was pushed up to $n=8$.
The results of these calculations can be summarized by the following tables,
which give the position of the FP and the eigenvalues $\lambda_i$
as functions of $n$.

\begin{center}
\centerline{\bf Position of Fixed Point ($\boldsymbol{\times10^{-3}}$)}
\begin{tabular}{crrrrrrrrr}
\toprule
 $n$ & $\tilde g_{0*}$ & $\tilde g_{1*}$ & $\tilde g_{2*}$ & $\tilde 
g_{3*}$
& $\tilde g_{4*}$ & $\tilde g_{5*}$ & $\tilde g_{6*}$& $\tilde g_{7*}$& $\tilde g_{8*}$\\
\midrule
1& 5.23& $-20.1$& & & & & & &\\
2& 3.29& $-12.7$& 1.51& & & & & &\\
3& 5.18& $-19.6$& 0.70& $-9.7\phantom{0}$& & & & &\\
4& 5.06& $-20.6$& 0.27& $-11.0\phantom{0}$& $-8.65$& & & &\\
5& 5.07& $-20.5$& 0.27& $-9.7\phantom{0}$& $-8.03$& $-3.35$& & &\\
6& 5.05& $-20.8$& 0.14& $-10.2\phantom{0}$& $-9.57$& $-3.59$& 2.46 & &\\
7& 5.04& $-20.8$& 0.03& $-9.78$& $-10.5\phantom{0}$ & $-6.05$& 3.42 & 
5.91 & \\
8& 5.07& $-20.7$& 0.09& $-8.58$& $-8.93$& $-6.81$ & 1.17 & 6.20 & 4.70\\
\bottomrule
\end{tabular}
\end{center}
%
%
%\smallskip
\begin{center}
\centerline{\bf Eigenvalues of linearized flow}
\begin{tabular}{crrrrrrrrr}
\toprule
 $n$ & $\operatorname{Re}\lambda_1$ & $\operatorname{Im}\lambda_1$ & 
$\lambda_2$ & $\lambda_3$
& $\operatorname{Re}\lambda_4$ & $\operatorname{Im}\lambda_4$ &  
$\lambda_6$ & $\lambda_7$& $\lambda_8$
\\
\midrule
1& $-2.38$& $-2.17$& & & & & & & \\
2& $-1.38$& $-2.32$& $-26.9$& & & & & &\\
3& $-2.71$& $-2.27$& $-2.07$& 4.23& & & & &\\
4& $-2.86$& $-2.45$& $-1.55$& 3.91& 5.22& & & & \\
5& $-2.53$& $-2.69$& $-1.78$& 4.36& 3.76 & 4.88 & & &\\
6& $-2.41$& $-2.42$& $-1.50$& 4.11& 4.42 & 5.98 & 8.58 & &\\
7& $-2.51$& $-2.44$& $-1.24$& 3.97& 4.57 & 4.93 & 7.57 & 11.1 &\\
8& $-2.41$& $-2.54$& $-1.40$& 4.17& 3.52 & 5.15 & 7.46 & 10.2 & 12.3\\
\bottomrule
\end{tabular}
\end{center}
From these numbers one can draw several conclusions.
First of all the FP exists for all truncations and secondly is relatively stable,
in the sense that adding new terms to the truncations generally does not
change very much the results of the lower truncation.
Third, there are three negative eigenvalues, showing that the critical
surface is three dimensional.
In fact, knowing the eigenvectors of the matrix $M$,
one can write explicitly the linearized equation of this surface.
Using $g_0$, $g_1$ and $g_2$ as independent parameters, 
\begin{align*}
&\tilde g_3=\phantom{-}0.00061243 + 0.06817374\,\tilde g_0 + 
0.46351960\,\tilde g_1 + 0.89500872\,\tilde g_2\\
&\tilde g_4=-0.00916502 - 0.83651466\,\tilde g_0 - 0.20894019\,\tilde g_1 + 1.62075130\,\tilde g_2\\
&\tilde g_5=-0.01569175 - 1.23487788\,\tilde g_0 - 0.72544946\,\tilde g_1 + 1.01749695\,\tilde g_2\\
&\tilde g_6=-0.01271954 - 0.62264827\,\tilde g_0 - 0.82401181\,\tilde g_1 - 0.64680416\,\tilde g_2\\
&\tilde g_7=-0.00083040 + 0.81387198\,\tilde g_0 - 0.14843134\,\tilde g_1 - 2.01811163\,\tilde g_2\\
&\tilde g_8= \phantom{-}0.00905830 + 1.25429854\,\tilde g_0 + 
0.50854002\,\tilde g_1 - 1.90116584\,\tilde g_2
\end{align*}
This illustrates the predictivity of asymptotically safe theories:
once the three parameters $g_0$, $g_1$ and $g_2$ have been measured at some scale
by means of three experiments,
everyting else is determined and any further experiment is a test of the theory.
Of course, the specific results of this calculation should not be taken too seriously:
there are many important things that have been neglected here.

%%%%%%%%%%%%%%%%%%%%%%%%%%%%%%%%%%%
\section{Matter}
%%%%%%%%%%%%%%%%%%%%%%%%%%%%%%%%%%%

However hard it may be to prove the asymptotic safety of gravity,
it would still not be enough: for applications to the real world 
one will have to show that a (possibly unified \cite{gravigut}) theory of all 
interactions is asymptotically safe.
The strong interactions are already described by an
asymptotically safe theory, and there are reasons to believe 
that this result is not ruined by the coupling to gravity \cite{flp}.
The electroweak and Higgs sectors of the standard model are perturbatively renormalizable,
but some of their beta functions are positive. This means that either new weakly coupled
degrees of freedom manifest themselves at some scale, before the couplings blow up,
or else the theory is consistent, but in a nonperturbative sense.
The simplest realization of the latter behavior is AS.
If the world is described by an AS theory, there are two main possibilities:
one is that AS is an inherently gravitational phenomenon,
in which case AS would manifest itself at the Planck scale
\footnote{this includes the possibility that due to the presence of 
large extra dimensions the effective Planck mass is much lower than $10^{19}$GeV.
I refer to \cite{litimplehn} for an analysis of this scenario.};
the other is that each interaction reaches the FP at its characteristic energy scale.

In the first case, one has to compute the effect of gravity on matter
couplings and the effect of matter on the gravitational couplings.
The effect of gravity on scalar couplings has been considered in \cite{griguolo,perini2,narain},
on gauge couplings in \cite{wilczek} and on Yukawa couplings in \cite{zzvp}.
One possibility is that the coupling to gravity
makes all matter interactions asymptotically free,
as conjectured long ago by Fradkin and Tseytlin \cite{FT}.
There is some evidence that this can happen in some cases,
with gravity preventing the Landau pole of scalar theory and QED
\cite{perini2, wilczek}.
In this case the second part of the job, namely computing the effect
of matter on gravity couplings, would be much simplified,
because in order to establish the existence of a FP it would be enough
to consider minimally coupled matter fields.
This problem has been studied in \cite{perini1},
where it was found that the existence of a FP with desirable properties puts 
restrictions on the number of matter fields of each spin.
In fact, for a large number of matter fields,
the task is even simplified, and to leading order in a $1/N$ expansion
one can prove the existence of a gravitational FP to all orders
of the derivative expansion \cite{largen}.
Things are more complicated if matter remains interacting also in the UV limit.
One particularly striking possibility has been pointed out recently \cite{harst}:
QED coupled to gravity seems to have two nontrivial FPs, in addition to the Gaussian one: at one gravity is interacting but QED is free, at the other they are both interacting. The latter has a lower dimensional critical surface and is therefore more predictive: on a renormalizable trajectory ending at this FP, the low energy value of the fine structure constant can in principle be calculated.

In the second case, matter and gravity would be separately AS.
Then, one would have to prove that electroweak theory somehow
heals itself of its UV problems.
At the moment, there are two approaches to this idea: 
the first, motivated by the formal analogies between gravity and the 
nonlinear sigma models, 
is that a Higgsless version of the standard model could be AS.
Some partial calculations support this view \cite{nlsm}. 
It has been shown recently that this possibility
is compatible with electroweak precision data \cite{fptv}. 
See also \cite{calmet} for comments. 
The other possibility is that a suitably balanced theory of coupled scalars and fermions
with potential and Yukawa couplings exhibits AS \cite{gies}.
In both cases the Higgs VEV, which is the source of the masses of all pointlike particles,
would run linearly above some scale, restoring scale invariance.
This would affect the physics of the Higgs, which is being explored at LHC,
making this by far the most exciting possibility from the point of view of
possible experimental signatures.

%%%%%%%%%%%%%%%%%%%%%%%%%%%%
\section{Cosmology and time}
%%%%%%%%%%%%%%%%%%%%%%%%%%%%

It is generally expected that a quantum theory of gravity
should be able to solve the puzzles that remain open in classical
general relativity, for example the fate of spacetime near a singularity.
Furthermore, a scale dependence of couplings (such as Newton's constant
or the cosmological constant) may well have an effect on the cosmological
evolution, even at relatively late stages.
For these reasons, cosmology, and especially very early cosmology,
is probably the most promising domain of application of asymptotically safe gravity.

The most popular way of applying the RG to cosmology consists 
in identifying the cutoff scale $k$
with some characteristic cosmological parameter (usually the Hubble scale $H(t)=\dot{a}(t)/a(t)$)
and then replacing the constant gravitational couplings ($G$, $\Lambda$...)
by their scale-dependent counterparts, making the gravitational couplings effectively time-dependent \cite{florean}.
This substitution can be done in a solution, in the equations of motion
or directly in the action, with different results.
Consider for example the effect of "RG-improving" Einstein's equations
\cite{entropysignature}:
\begin{equation}
G_{\mu\nu}=8\pi G(k)T_{\mu\nu}-\Lambda(k) g_{\mu\nu}\ .
\end{equation}
For simplicity we assume a spatially flat Friedmann-Robertson-Walker metric
with scale factor $a(t)$ and
an energy momentum tensor in the form of a perfect fluid
$T^{\mu}_{\phantom{\mu}\nu}=diag(-\rho,\,p,\,p,\,p)$ 
with equation of state $p(\rho)=w\rho$.
Both $G_{\mu\nu}$ and $R_{\mu\nu}$ can be expressed
in terms of the Hubble rate:
$$R_{tt}=-3\,(\dot{H}+H^2)\ ,\qquad 
R=R^{\mu}_{\phantom{\mu}\mu}=6(\dot{H}+2H^2)\ ,\qquad
G_{tt}  =  3H^2\ ,
$$
so that the $(tt)$-component and the trace of Einstein's equations become
\begin{eqnarray}
\label{e1}
3H^2& = & 8\pi G\rho+\Lambda\ ,\\
\label{e2}
6(\dot H+2H^2)& = & 8\pi G\rho\,(1-3w)+4\Lambda\ .
\end{eqnarray}
Choose the cutoff $k=\xi H$, for some real number $\xi$ of order one.
Then Newton's constant and the cosmological constant become functions of time:
$G=G(\xi H)$, $\Lambda=\Lambda(\xi H)$, whose form is fixed by the
renormalization group equations.
To simplify, let us assume that we are at sufficiently high $k$ such
that we may assume that the (dimensionless) couplings $\tilde G$ and
$\tilde\Lambda$ are at their fixed point values.
Then $G=\tilde G_*/(\xi^2 H^2)$ and $\Lambda=\tilde\Lambda_*\xi^2 H^2$.
One then looks for inflationary de Sitter solutions
\begin{equation}
\label{exponential}
a(t)=a_0 e^{Ht}\ ;\qquad
H=\mathrm{constant}\ ,
\end{equation}
or power law solutions
\begin{equation}
\label{powerlaw}
a(t)=a_0 t^p\ ;\qquad
H=\frac{p}{t}\ .
\end{equation}
The equations admit power law solutions with 
\begin{equation}
\label{solp1less}
p=\frac{2}{(3-\tilde\Lambda_*\xi^2)(1+w)}\ .
\end{equation}
Let us set $w=1/3$, as appropriate for ultrarelativistic matter.
We see that for $1/2<\tilde\Lambda_*\xi^2/3<1$
the solution has inflationary character ($p>1$),
with the acceleration becoming stronger as $\tilde\Lambda_*\xi^2/3$ increases.
For $\tilde\Lambda_*\xi^2/3=1$ (and any $w>-1$) the exponent diverges.
We observe that this condition is equivalent to
the equation $R=4\Lambda$ written in the FP regime;
the corresponding solution is a de Sitter universe.
Similar conclusions have been shown to hold also for the fixed point
of $f(R)$ gravity \cite{bcp}, and first steps towards a calculation
of the spectrum of fluctuations have been made in \cite{contillo}.
A general qualitative analysis of the cosmological dynamics in the presence of
running couplings has been given recently in \cite{hindmarsh}.

This approach raises several issues.
One is that inflation is supposed to occur at energies considerably lower than the
Planck scale, so that the approximation of being close to the fixed point
may actually not be warranted \cite{tye}.
Another issue is the exit from inflation. Presumably this would happen when the
RG trajectory departs from the immediate neighborhood of the fixed point,
but a detailed study has not been done so far.
Perhaps more worrisome is the nonconservation of the matter energy-momentum tensor.
From Friedmann's equations one obtains a modified conservation equation
\begin{equation}
\nonumber
\dot\rho+3H(\rho+p)=
-\frac{1}{8\pi G}(\dot\Lambda+8\pi\rho \dot G)
\end{equation}
We see that the time variation of the couplings,
which follows from the time dependence of the cutoff,
gives rise to nonconservation of the energy.
One may try to interpret this in terms of the energy and momentum of the
field modes that have been removed from the system by coarse graining.
Bonanno and Reuter actually turn this into a positive feature \cite{entropysignature}:
they show that, under reasonable assumptions, the energy transferred 
to the matter system through
the decay of the cosmological constant over the age of the universe 
is of the correct order of magnitude
to explain the entropy of the cosmic background radiation.

In order to avoid these issues, Weinberg follows a different approach
\cite{weinberginflation}.
He writes the Friedmann equations following from the most general
effective action that is local in curvatures and covariant derivatives of curvatures,
and looks for de Sitter solutions.
He argues that a choice of cutoff of the order of $H$ may be a reasonable
compromise between the conflicting requirements of avoiding large radiative corrections
to the field equations, and the Einstein-Hilbert truncation being a resonable approximation.
In this approach the exit from inflation should be signalled by an instability 
of the solution.
Unfortunately explicit calculations based on known properties of
the fixed point of pure gravity seem to show too much instability,
leading to a number of $e$-foldings that is too small.

Aside from these attempts to apply asymptotic safety to inflationary
cosmology, one may try to make connection also to other ideas.
One important fact is  that physics at a fixed point is scale invariant
\footnote{due to the complex critical exponents, one may only have
invariance under a discrete subgroup of scale transformations.}.
Even though the fixed point Lagrangian contains dimensionful couplings, these
scale with energy according to their canonical dimension
so that all observable quantities have power law dependences.
Under these circumstances, defining a clock becomes impossible
even in principle and the notion of time loses its operational meaning \cite{zinkernagel}.
Although one may still be able to define separate points, time intervals
and distances become meaningless.
In this sense, one may argue that a fixed point leads to a notion
of minimal distance \cite{minl}.
This is also in line with the view that the metric geometry
``melts down'' near the big bang, but the conformal geometry remains well defined.
In fact it is worth noting that if the infrared behavior of gravity was
also governed by a fixed point, as conjectured in \cite{benti},
then one would have scale invariance at both ends of the cosmological evolution.
This would lend support to Penrose's Conformal Cyclic Cosmology \cite{penrose}.

%%%%%%%%%%%%%%%%%%%%%%%%%%%%%%%%%%%%%%%%%%%
\section{Discussion, summary and prospects}
%%%%%%%%%%%%%%%%%%%%%%%%%%%%%%%%%%%%%%%%%%%

I have presented some evidence that a theory of gravity 
and perhaps of all interactions is AS. 
None of the calculations performed so far can be said to be a proof,
but the qualitative agreement of the results in all the approximations 
makes this by now a rather plausible scenario.
If this was true, we would have an UV complete theory remaining within 
the familiar domain of QFT.
It is important to appreciate the differences between this and other popular 
approaches to quantum gravity.

AS is a ``bottom up'' approach to quantum gravity:
the discussion starts within the theory space of an effective field theory,
and goes on to note that if the world corresponds to a trajectory of a special type,
then the effective description can be pushed to arbitrarily high energy.
An AS theory is simply the continuation of an effective theory to higher energy scales.
As a result, an AS theory has the great advantage that if it exists,
it is almost automatically in agreement with our knowledge of the low energy world.
This is in contrast to string theory and loop quantum gravity, which are ``top down'' approaches. For them, making a connection with known low energy phenomenology 
is proving a very hard issue.

There is obviously a price to pay for this. On one hand, in a nonperturbative context
it is hard to obtain reliable results and hard proofs.
Furthermore, the action of the FP theory seems to contain infinitely many terms
with nonzero couplings, making it unwieldy at best.
It is in principle possible that the description of the fixed point could 
be simplified by a suitable change of variables (perhaps along the lines of \cite{krasnov}). 
Then, the AS QFT may turn out to be equivalent to one of the top down theories.
In that case it would be enough to establish the equivalence in the vicinity of the Planck scale. From there downwards, one would just follow the RG as in any effective field theory.

This remark applies also to the scenario of ``emergent gravity''.
According to a popular point of view, gravity is not a fundamental interaction
but rather the effective description of some underlying microscopic dynamics
that may have little to do with the geometry
\cite{adler, akama, emergent}.
It is often said that in this case attempts at formulating a quantum theory of gravity 
in terms of metric degrees of freedom are misplaced.
As discussed in \cite{vacca}, even if gravity at very high energies
was described by some as yet unknown theory with non-metric degrees of freedom,
from some energy scale downwards it can be described by an effective theory of the metric,
and in this effective theory couplings will run according to the RG as discussed above.
At sufficiently low energy we would therefore be again in the theory space
discussed in section 2.
If there is a FP in this theory space, then the RG trajectory that describes
emergent gravity will approach its UV critical surface at low energies,
so that even in this case the notion of AS would prove to be a useful tool.

\cleardoublepage

\end{document}